\documentclass[pra, twocolumn,showpacs,amsmath,amssymb]{revtex4}

\usepackage{graphicx}%Include figure files
\usepackage{dcolumn}%Align table columns on decimal point
\usepackage{bm}% bold math

\begin{document}
\title{Development of a configuration-interaction + all-order method for atomic calculations}

\author{M. S. Safronova}
 \homepage{http://www.udel.edu/~msafrono}
 \email{msafrono@udel.edu}
\affiliation{Department of Physics and Astronomy, University of
Delaware, Newark, DE 19716-2570, USA}

\author{M. G. Kozlov}
\affiliation{Petersburg Nuclear Physics Institute, Gatchina 188300, Russia}

\author{W. R. Johnson}
\affiliation{Department of Physics, 225 Nieuwland Science Hall,
University of Notre Dame, Notre Dame, Indiana, 46556, USA}

\author{Dansha Jiang}
\affiliation{Department of Physics and Astronomy, University of
Delaware, Newark, DE 19716, USA}

\begin{abstract}
We develop a theoretical method within the framework of relativistic
many-body theory to accurately treat correlation corrections in
atoms with few valence electrons. This method combines the all-order
approach currently used in precision calculations of properties of
monovalent atoms with the configuration-interaction approach that is
applicable for many-electron systems. The method is applied to
Mg, Ca, Sr, Zn, Cd, Ba, and Hg to evaluate ionization energies and
low-lying energy levels.
\end{abstract}

\maketitle

\section{Introduction}
The development of the relativistic all-order method where all single and double excitations
of the Dirac-Hartree-Fock wave function are included to all orders of perturbation
theory led to accurate predictions for energies, transition amplitudes,
hyperfine constants, polarizabilities, $C_3$
and $C_6$ coefficients, isotope shifts,
and other properties of monovalent atoms, as well as the calculation of parity-violating
amplitudes in Cs, Fr, and Ra$^+$ \cite{SD,SDpT,review07,pnc-cs,pnc-fr,pnc-ra}.
This method was also used to calculate magic  wavelengths \cite{magic} as well as
black-body radiation shifts \cite{BBR} and quadrupole moments \cite{quad}
that are of interest to atomic-clock research.
 The all-order method is designed to treat core-core and core-valence
correlations with high accuracy.
 It is one of the most accurate methods currently being used in the atomic structure
calculation. However, its applications so far have been limited to
monovalent systems. Readers are referred to
Ref.~\cite{review07} and references therein for a review of this
method and its applications.

Precision calculations for atoms with several valence electrons
require an accurate treatment of the very strong valence-valence
correlation; a perturbative approach leads to significant
difficulties. The complexity of the all-order formalism for matrix
elements also increases drastically as the number of valence
electrons increases; for example, the expression for all-order
matrix elements in divalent systems contains several hundred terms
instead of the twenty terms in the corresponding monovalent
expression. Therefore, we found it impractical to develop a direct
extension of the all-order approach to more complex systems, both
due to the large valence-valence correlation corrections and
the very large
number of terms noted above in the matrix element formulas.

A more promising method for the study of atomic properties of
more complicated systems that combined configuration interaction
(CI) and many-body perturbation theory (MBPT) was developed in
Ref.~\cite{Dzuba:1996}. The CI+MBPT method was applied to the
calculation of atomic properties of various systems in a number of
works (see
\cite{Porsev:2001,Kozlov:1999,Kozlov:1997,Dzuba:2007,Dzuba:2007a,Savukov:2002,Savukov:2002a,Savukov:2003,DF1,DF2}
and references therein) and to the calculation of PNC amplitudes in
Tl and Yb \cite{KPJ:01,PKR00}. The strengths of the
configuration-interaction method are broad applicability and
all-order treatment of the valence-valence correlation corrections.
However, the precision of the CI method is generally
drastically limited for large systems by the number of the
configurations that can be included. As a result, core excitations
are neglected or only a small number of them are included, leading
to a significant loss of accuracy for heavier systems. The CI+MBPT
approach allows one to incorporate core excitations in the CI method
by constructing an effective Hamiltonian $H^\text{eff}$ that
incorporates certain perturbation theory terms. The CI method is
then applied to the modified $H^\text{eff}$ to obtain improved
energies and wave functions.

Because of the rapid increase in the number of terms of the MBPT
expansion, the CI + MBPT approach becomes impractical already in the
third order of MBPT. For that reason, the CI + MBPT approach is usually restricted to
second order. Some higher-order corrections can be accounted
for by introducing screening coefficients to second order
diagrams. These screening coefficients can be found either by
averaging two-electron second order diagrams, or by semi-empirical
fitting of experimental energies.
The second-order expression for one-body correction to the Hamiltonian
is corrected by the all-order chains of such terms in some works (see, for example,
Ref.~\cite{Dzuba:2006}).
 In 2004, a modification of the
effective Hamiltonian using the all-order pair equations was
proposed and tested on a ``toy'' 4-electron model \cite{Kozlov:all}.
% >>MSS
An efficient method of including core-valence correlations into the configuration
interaction (CI) calculations was presented by Dzuba and Flambaum in \cite{DF:2007}.
The CI Hamiltonian for N valence electrons
was calculated using orbitals in the complete $V^N$ potential (the mean field produced by all electrons);
the one- and two-body corrections to the effective Hamiltonian were obtained by
 using many-body perturbation theory with  dominating
classes of diagrams included to all orders.

In the present work, we  combine the all-order method, currently used in
precision calculations of properties of monovalent atom, with the
configuration interaction (CI) approach. In the CI + all-order
approach, core excitations are incorporated in the CI method by
constructing an effective Hamiltonian using fully converged
all-order excitation coefficients. Therefore, the core-core and
core-valence sectors of the correlation corrections for systems with
few valence electrons will be treated with the same accuracy as in
the all-order approach for monovalent atoms. Then, the CI method
is used to treat valence-valence correlations. This method is
expected to yield accurate wave functions for subsequent
calculations of atomic properties such as lifetimes,
polarizabilities, hyperfine constants, parity-violating amplitudes,
etc. The present work is motivated by the urgent need for precision calculations
of atomic properties of heavy atoms with few valence electrons for
applications such as atomic clock research, quantum
information, study of fundamental symmetries, searches for variation
of the fundamental constants,  and tests of high-precision
experimental methods. The development of the CI + all-order  method
is also aimed at filling the long-standing gap between the accuracy
of theoretical and  experimental parity-violation studies in
systems with few valence electrons. Atomic properties of  various
atoms and ions are also of interest for astrophysics applications.

Our method is generally applicable, i.e. not restricted to the
specific type of the system. We test the present approach on the
calculation of the energy levels of Mg, Ca, Sr, Cd, Zn, Ba, and Hg to
demonstrate a significant improvement in comparison with CI + second-order
MBPT values.  We also discuss calculations of
transition matrix elements and polarizabilities.

We provide a brief description of the all-order and CI +
MBPT formalisms in Sections \ref{all} and \ref{MBPT}, respectively.
The CI + all-order approach is described in Section~\ref{CI-all}.
Finally, we present results for removal energies in divalent
systems calculated using the CI + all-order approach and discuss
perspectives for further applications and development.

\section{Relativistic all-order method for monovalent systems}
\label{all}

Our point of departure is the relativistic no-pair Hamiltonian $H=H_0+V_I$
obtained from QED by \citet{BR:51}:
\begin{align}
H_0 & = \sum_i \epsilon_i :a_i^\dagger a_i:\, , \label{np1}\\
V_I & = \frac{1}{2} \sum_{ijkl} v_{ijkl}  : a^\dagger_ia^\dagger_j a_l a_k :
  + \sum_{ij} \left(V_\text{HF} - U\right)_{ij} :a^\dagger_ia_j :  \, . \label{np2}
\end{align}
Here,  $v_{ijkl}$ are two-particle matrix elements of the Coulomb
interaction $g_{ijkl}$, or Coulomb + Breit interaction
$g_{ijkl}+b_{ijkl}$, and $V_{HF}=\sum_{a}\left( v_{iaja} -
v_{iaaj}\right)$ is \textit{frozen-core} Dirac-Fock potential. The
summation index $a$ in $V_{HF}$ ranges over  states in the closed
core. The quantity $\epsilon_i$ in Eq.~(\ref{np1}) is the eigenvalue
of the Dirac equation $h(\bm{r}) \phi_i(\bm{r}) = \epsilon_i
\phi_i({\bm r}), $ where
\begin{equation}
h(\bm{r}) = c\, {\bm \alpha}\cdot {\bm p} + \beta m c^2 -\frac{Z}{r} + U(r)\, . \label{dh:2}
\end{equation}
In our previous all-order calculations of monovalent atoms, we took
$U$ to be \textit{frozen-core} $V^{N-1}$ potential, $U=V_{HF}$. Such
a choice greatly simplifies the calculations since the
second term in Eq.~(\ref{np2}) vanishes in this case. In this work,
we use the same type of potential ($V^{N-2}$ for divalent systems) ,
but different potentials may be used in the future.

In the coupled-cluster method, the exact many-body wave function is
represented in the form \cite{CK:60}
\begin{equation}
|\Psi \rangle = \exp(S) |\Psi^{(0)}\rangle, \label{cc}
\end{equation}
where $|\Psi^{(0)}\rangle$ is the lowest-order atomic state vector.
The operator $S$ for an N-electron atom consists of
``cluster'' contributions from one-electron, two-electron, $\cdots$,
N-electron excitations of the lowest-order state vector $|\Psi^{(0)}\rangle$:
$S=S_1+S_2+ \dots +S_N$.

The all-order method described in detail in Refs.~\cite{SD,SDpT}, is a
linearized version of the coupled-cluster method, where all non-linear terms
in the expansion of the exponential are omitted; the all-order wave function takes the form
\begin{equation}
|\Psi \rangle = \left\{1+S_1+S_2+S_3 + \cdots +S_N \right\}
|\Psi^{(0)}\rangle\, . \label{cc2}
\end{equation}

Restricting the sum in Eq.~(\ref{cc2}) to single, double, and valence triple
excitations yields the following expansion for the state vector
of a monovalent atom in state $v$:
\begin{eqnarray}
\lefteqn{ |\Psi_v \rangle = \left[ 1 + \sum_{ma} \, \rho_{ma}
a^\dagger_m a_a + \frac{1}{2} \sum_{mnab} \rho_{mnab} a^\dagger_m
a^\dagger_n a_b a_a +
 \right. } \hspace{0.5in} \label{eq1} \\
&& + \sum_{m \neq v} \rho_{mv} a^\dagger_m a_v + \sum_{mna}
\rho_{mnva} a^\dagger_m a^\dagger_n a_a a_v
\hspace{0.5in} \nonumber \\
&& + \left.
\frac{1}{6}\sum_{mnrab}
\rho_{mnrvab} a^\dagger_m a^\dagger_n a^\dagger_r a_b a_a a_v
\right]| \Psi_v^{(0)}\rangle\nonumber
,
\end{eqnarray}
where the indices $m$, $n$, and $r$ range over all possible virtual
states while indices $a$ and $b$ range over all occupied core
states. The lowest-order
 wave function $| \Psi_v^{(0)}\rangle$  is
 \begin{equation}
 | \Psi_v^{(0)}\rangle = a^\dagger_v | \Psi_{C}\rangle,
  \end{equation}
where $| \Psi_{C}\rangle$ is the lowest-order frozen-core wave
function. The quantities $\rho_{ma}$, $\rho_{mv}$ are
single-excitation coefficients for core and valence electrons;
$\rho_{mnab}$ and $\rho_{mnva}$ are core and valence
double-excitation coefficients, respectively; $\rho_{mnrvab}$ are
the valence triple excitation coefficients. In the single-double (SD)
implementation of the all-order method, only single and double
excitations are included. In the (single, double, partial triple)
SDpT variant of the all-order method, valence triple excitations are
included perturbatively as described in Ref.~\cite{SDpT}.

To derive equations for the excitation coefficients, the state
vector $|\Psi_v\rangle$ is substituted into the  many-body
Schr\"{o}dinger equation $H | \Psi_v\rangle=E| \Psi_v\rangle,
\label{eq2}$ and terms on the left- and right-hand sides are
matched, based on the number and type of operators they contain,
leading to the  equations for the  excitation coefficients given in
\citet{SDpT}.

We note that all non-linear terms at the single-double level have been
added in the formulation of the all-order method in Ref.~\cite{NL}.
This version of the all-order method is equivalent to the
coupled-cluster single-double (CCSD) method for a finite basis set.
It was shown in Refs.~\cite{AS-1,Andrei:Cs} that both non-linear
terms and complete valence triple excitations must be included to
improve the accuracy of the linearized coupled-cluster SD method. In
the present work, we use linearized SD variant of the all-order method
since the SD method already leads to excellent results for a large number of
the atomic properties and is computationally efficient.

The resulting SD all-order equations for valence excitation coefficients are:
\begin{eqnarray}
&&(\epsilon_v-\epsilon_m + \delta E_v)\rho_{mv} =
\label{v1}  \\
&& \sum_{bn} \tilde{g}%
_{mbvn}\rho_{nb} + \sum_{bnr} g_{mbnr}\tilde{\rho}_{nrvb} - \sum_{bcn}
g_{bcvn}\tilde{\rho}_{mnbc}\,\nonumber ,
\end{eqnarray}
\begin{eqnarray}\label{v2}
&&(\epsilon_{vb}-\epsilon_{mn} + \delta E_v)\rho_{mnvb} =  \\
&& g_{mnvb} +
\sum_{cd} g_{cdvb}\rho_{mncd} + \sum_{rs} g_{mnrs}\rho_{rsvb} \hspace{0.25in}
\notag \\
&& + \left[ \sum_{r} g_{mnrb}\rho_{rv} -\sum_{c} g_{cnvb}\rho_{mc} +
\sum_{rc} \tilde{g}_{cnrb}\tilde{\rho}_{mrvc} \right] \nonumber \\&& + \left[
\begin{array}{c}
v \leftrightarrow b \\
m \leftrightarrow n%
\end{array}
\right]\,\nonumber ,
\end{eqnarray}
where $\delta E_v$ is the valence correlation energy $\delta E_v= E_v-\epsilon_v$, $\epsilon_{ij}=\epsilon_i+%
\epsilon_j$, and  $\tilde{\rho}%
_{mnvb} ={\rho}_{mnvb} -{\rho}_{nmvb}$. The correlation correction to the energy
of the state $v$, is given in terms of the excitation coefficients by
\begin{equation}
\delta E_v = \sum_{ma} \tilde{g}_{vavm}\rho_{ma} + \sum_{mab} g_{abvm}\tilde{%
\rho}_{mvab} + \sum_{mna} g_{vbmn}\tilde{\rho}_{mnvb}.  \label{env}
\end{equation}
Equations for core excitation coefficients $\rho_{ma}$ and
$\rho_{mnab}$ are obtained from the above equations by replacing the
valence index $v$ by a core index $a$ and removing $\delta E_v$ from
the left-hand side of the equations. We note that the  right-hand side of
the valence energy equation is identical to the right-hand side of the
equation for $\rho_{mv}$
with $m=v$. Equations for the correlation energy and all
excitation coefficients are solved iteratively. Every iteration
picks up correlation terms corresponding to the next higher order of
perturbation theory  until the correlation energy converges to
sufficient numerical accuracy. Therefore, the all-order approach
includes dominant MBPT terms to all-orders.

Matrix elements for any one-body operator $Z = \sum_{ij} z_{ij}\
a^\dagger_i a_j$ are obtained within the framework of the all-order
method as
\begin{equation}
Z_{wv}=\frac{\langle \Psi_w |Z| \Psi_v \rangle}{\sqrt{\langle \Psi_v |
\Psi_v \rangle \langle \Psi_w | \Psi_w \rangle}},  \label{eqr}
\end{equation}
where $|\Psi_v\rangle$ and $|\Psi_w\rangle$ are given by the
expansion (\ref {eq1}). In the SD approximation, the resulting
expression for the numerator of Eq.~(\ref{eqr}) consists of the sum
of the DF matrix element $z_{wv}$ and 20 other terms that are linear
or quadratic functions of the excitation coefficients. The advantage
of this approach is that the expression in Eq.~(\ref{eqr}) does not
depend on the nature of the operator $Z$, only on its rank and
parity. Therefore, matrix elements of any one-body operator may be
calculated with the same general code.

The complexity of the all-order formalism for matrix elements
increases drastically with the number of valence electrons.
We have derived the expression for all-order matrix elements in
divalent systems; it contains several hundred terms instead of the
twenty terms in the corresponding monovalent expression. Therefore,
it is impractical to extend the all-order method to
the case of more complicated atoms in its present implementation directly,
i.e.\ to start single-double expansion from the divalent
lowest-order wave function containing $a^\dagger_w
a^\dagger_v|\Psi_{C}\rangle$. Such an approach
also leads to ``intruder state'' problems
well-known in the perturbation expansions
based on the Rayleigh-Schr\"{o}dinger implementation of the MBPT.

We note that the relativistic couple-cluster method has been
successfully applied to calculation of the energies and electron
affinities in systems with few electrons (see
Refs.\cite{CCSD-1,CCSD-2,CCSD-3,CCSD-4,CCSD-5} and references
therein).
%%mgk>
It is rather difficult to apply this method for calculation of other
atomic properties, such as transition matrix elements. However,
using  the \textit{finite field} technique one can use this method,
for example, to calculate quadrupole hyperfine constants in such a
heavy atoms as Au \cite{YEK07}.

\section{CI + MBPT method}
\label{MBPT}

A combination of the configuration-interaction (CI) method and
perturbation theory was developed in Ref.~\cite{Dzuba:1996}.
%%mgk>>>
It was based on the Dirac-Fock code \cite{BDT77} and CI code
\cite{KT87}.
%%mgk<<
This approach has been later applied to the calculation of atomic
properties of various systems in a number of works
(see %%@
\cite{Porsev:2001,Kozlov:1999,Kozlov:1997,Dzuba:2007,Dzuba:2007a,Savukov:2002,Savukov:2002a,Savukov:2003} %%@
and references therein).
%%mgk>>
In Refs.\ \cite{KPJ:01,PKR00} the CI + MBPT method was used to calculate PNC
amplitudes in Tl and Yb respectively (in the latter case the
nuclear-spin-dependent amplitude was calculated).

In the CI method, the many-electron wave function is obtained as a
linear combination of all distinct states of a given angular
momentum J and parity \cite{Savukov:2002}:
\begin{equation}
\Psi_{J} = \sum_{i} c_{i} \Phi_i,
\end{equation}
in other words, a linear combination of Slater determinants of
proper symmetry from a model subspace \cite{Dzuba:1996}.

Energies and wave functions of low-lying states are determined by
diagonalizing an effective Hamiltonian:
\begin{equation}
H^{\text{eff}}=H_{1} + H_2, \label{ham}
\end{equation}
where $H_{1}$ represents the one-body part of the Hamiltonian, and
$H_2$ represents the two-body part (Coulomb or Coulomb + Breit matrix
elements $v_{ijkl}$). We use Coulomb matrix elements $g_{ijkl}$ in
the present work. The resulting wave functions are used to calculate
matrix elements and other properties such as polarizabilities,
parity-violating amplitude, etc. The precision of the
configuration-interaction method is drastically limited for large
systems by the number of the configurations that can be included.
Consequently, core excitations are entirely omitted or only a small number
are included, leading to a significant loss of accuracy for
heavy atoms.

The CI + MBPT approach allows one to incorporate core excitations in
the CI method by including certain higher-order terms in an
effective Hamiltonian~(\ref{ham}). The CI method is then applied as
usual with the modified $H^\text{eff}$ to obtain improved energies
and wave functions. Somewhat different versions of the CI+MBPT
method exist; here, we describe the approach used in the present
work and follow the designations of Ref.~\cite{Dzuba:2006}.

In the CI+MBPT approach, the one-body part $H_1$ is modified to include
the correlation potential $\Sigma_1$ that accounts for part of the core-valence correlations,
\begin{equation}\label{H1eff}
H_1 \rightarrow H_1+\Sigma_1.
\end{equation}
 Either the second-order expression, $\Sigma_1^{(2)}$, or
all-order chains of such terms can be used (see, for example, %%@
Ref.~\cite{Dzuba:2006}).
The latter approach corresponds to replacing Dirac-Fock orbitals by Brueckner orbitals.
The second-order matrix elements $(\Sigma^{(2)}_1)_{yx}$ are given by
\begin{equation}
\label{11}
\left(\Sigma^{(2)}_1\right)_{yx}=\sum_{mab}\frac{g_{myab}\,\tilde{g}_{mxab}}{\epsilon_{ab}-\epsilon_{xm}+\tilde{\epsilon}_y-\epsilon_y}
+
\sum_{mna}\frac{g_{mnxa}\,\tilde{g}_{mnya}}{\tilde\epsilon_{y}+\epsilon_{a}-\epsilon_{mn}}.
\end{equation}
We use the same designations as in Section~\ref{all}; indices
from the middle of the alphabet $m$ and $n$ range over all possible
virtual states while indices $a$ and $b$ range over all occupied
core states. The one-particle energies $\epsilon_i$ are written
together as $\epsilon_{ij}=\epsilon_i+\epsilon_j$ for brevity. The
summation over index $i$ implies the sum over the quantum numbers
$n_i\,\kappa_i\,m_i$.

The CI+MBPT approach is based on the Brilloiun-Wigner variant of
MBPT, rather than the Rayleigh-Schr\"{o}dinger variant. Use of the
Rayleigh-Schr\"{o}dinger MBPT for systems with more than one valence
electron leads to a non-symmetric effective Hamiltonian and to the
problem of ``intruder states''. In the Brilloiun-Wigner variant,
 the effective Hamiltonian is symmetric and accidentally small
denominators do not arise; however,  $\Sigma_1$ and $\Sigma_2$
became energy dependent. Specifically, the one-body correction
$\Sigma_1$ depends on the energy $\tilde\epsilon_y$ (see
Eq.~(\ref{11})).
Ideally, the energy $\tilde{\epsilon}_y$ should be calculated from
the particular eigenvalue of the effective Hamiltonian \eqref{ham}.
In practice, we use several approaches. The simplest and the most
practical one is to set the energy $\tilde\epsilon_{y}$ to the
Dirac-Fock energy of the lowest orbital for the particular partial
wave. For example, we use $\tilde\epsilon_{ns}=\epsilon_{3s}$  for
all $ns$ orbitals of Mg system. This approximation usually works
reasonably well for atomic states belonging to the lowest
configurations of a given symmetry.
Another approach is to set the energy of all orbitals for
a particular partial wave to a certain value, one value is specified for
each partial wave. This approach allows one to generate better atomic wave functions
for subsequent evaluation of the atomic properties by
selecting the values of $\tilde\epsilon$ so final energy eigenstates are tuned to the
experimental values.
%%mgk<<
We have also developed a more elaborate method that involves
calculating derivatives of the $\Sigma_{1}$ and $\Sigma_{2}$ with
respect to $\tilde\epsilon$ that allows to adjust the effective
Hamiltonian as suggested in \cite{Dzuba:1996}.
Our implementation of the CI+all-order method permits us to use any
of these strategies.

  Performing analytical sums over all magnetic quantum numbers yields the expression
  \begin{eqnarray}
\left(\Sigma^{(2)}_1\right)_{yx}=&-&\sum_{mab}\sum_{K}\frac{1}{\left[K\right]\left[j_y\right]}
\frac{X_K(myab)Z_K(mxab)}{\epsilon_{ab}-\epsilon_{xm}+\tilde{\epsilon}_y-\epsilon_y} \hspace{0.4in} \nonumber \\
&+&
\sum_{mna}\sum_K\frac{1}{\left[K\right]\left[j_y\right]}\frac{X_K(mnxa)
Z_K(mnya)}{\tilde\epsilon_{y}+\epsilon_{a}-\epsilon_{mn}},
\end{eqnarray}
where $K$ is multipolarity restricted by conventional triangular
rules, $\left[ K\right] =2K+1$, and the summations over all
lower-case indexes $i$ now
 designate sums over $n_i$ and $\kappa_i$. We use similar designations for the sums
 listed below
to avoid explicitly writing out all quantum numbers in all sums.
  The quantity $X_K(mnab)$ is
\begin{eqnarray}
&&X_K(mnab)= \\
&&(-1)^K\left\langle \kappa _m\left\| C^K\right\| \kappa
_a\right\rangle \left\langle \kappa _n\left\| C^K\right\| \kappa
_b\right\rangle R_K(mnab)\nonumber,
\end{eqnarray}
where $R_K(mnab)$ are (relativistic) Slater integrals and
$\left\langle \kappa _m\left\| C^K\right\| \kappa _a\right\rangle $
are reduced matrix elements of a normalized spherical harmonics.
$Z_K(mnab)$ is given by
\begin{eqnarray}
Z_K(mnab) &=& X_K(mnab)  \\&+& \sum_{K'} \left[K\right] \left\{ \begin{array}{ccc}
 j_m  & j_a & K \\
 j_n  & j_b & K'
 \end{array} \right\} X_{K'}(mnba) \nonumber.
\end{eqnarray}

The two-body Coulomb interaction term $H_{2}$ is modified by including
the two-body part of the core-valence interaction that represents
screening of the Coulomb interaction by valence electrons;
\begin{equation}\label{H2eff}
H_2 \rightarrow H_2+\Sigma_2,
\end{equation}
 where $\Sigma_2$ is calculated in second-order MBPT in CI+MBPT approach.
The second-order matrix elements $(\Sigma^{(2)}_2)_{mnvw}$ are given by
\begin{eqnarray}
&&\left(\Sigma^{(2)}_2\right)_{mnvw}=\sum_{cd}\frac{g_{vwcd}\,{g}_{mncd}}{\epsilon_{cd}-\epsilon_{mn}+\tilde{\epsilon}_v-\epsilon_v
+\tilde{\epsilon}_w-\epsilon_w} \hspace{0.3in}\\
&&+\left[ \sum_{rc}\frac{\tilde{g}_{wrnc}\,\tilde{g}_{mrvc}}{\tilde\epsilon_{v}+\epsilon_{c}-\epsilon_{mr}+\tilde{\epsilon}_w-\epsilon_w}
+ \left( \begin{array}{ccc}
 m  & \Leftrightarrow & n \\
 v  &  \Leftrightarrow& w
 \end{array} \right)\right].\nonumber
\end{eqnarray}
Performing an angular reduction leads to
\begin{eqnarray}
&&(\Sigma^{(2)}_2)_K(mnvw)=\\
&&\sum_{cd}\sum_{LK'}[K]
\left\{ \begin{array}{ccc}
 j_m  & j_v & K \\
 L  & K' & j_c
 \end{array} \right\}
\left\{ \begin{array}{ccc}
 j_n  & j_w & K \\
 L  & K' & j_d
 \end{array} \right\} \nonumber \\
&&\times
\frac{X_L(vwcd)\,X_{K'}(cdmn)}{\epsilon_{cd}-\epsilon_{mn}+\tilde{\epsilon}_v-\epsilon_v
+\tilde{\epsilon}_w-\epsilon_w} \nonumber \\
&&- \sum_{rc} \frac{(-1)^{j_w+j_n+K}}{[K]}\frac{Z_K(wrnc)\,Z_K(mrvc)}{\tilde\epsilon_{v}+\epsilon_{c}-\epsilon_{mr}+\tilde{\epsilon}_w-\epsilon_w} \nonumber\\
&&- \sum_{rc}\frac{(-1)^{j_v+j_m+K}}{[K]}\frac{Z_K(vrmc)\,Z_K(nrwc)}{\tilde\epsilon_{w}+\epsilon_{c}-\epsilon_{nr}+\tilde{\epsilon}_v-\epsilon_v}. \nonumber
\end{eqnarray}
MBPT corrections associated with terms $\Sigma_1$ in Eq.\
\eqref{H1eff} and $\Sigma_2$ in Eq.\ \eqref{H2eff} typically grow
with nuclear charge $Z$ leading to a deterioration of the accuracy
of the CI + second-order MBPT results for heavier, more complicated
systems. The order-by-order extension of this method does not look
promising for two reasons. First, the complexity of the MBPT
expansion for systems with more than one valence electron already makes
third-order calculations impractical. Second, the convergence of the MBPT
series is not well studied, but it is known that third order
is often less accurate than second order. This is why it
was so important to develop an all-order extension of the MBPT
method for monovalent systems.
%%mgk<<

\section{CI + all-order method}
\label{CI-all}

  In the CI + all-order approach, corrections to the effective Hamiltonian
  $\Sigma_1$ and $\Sigma_2$ are calculated using the all-order method,
  in which the effective Hamiltonian
  contains dominant core and core-valence correlation corrections to all orders,
   as discussed in Section~\ref{all}.  The core-core and
  core-valence sectors of the correlation corrections for systems with few valence electrons
  are treated in the all-order method with the same accuracy
   as in the all-order approach for the monovalent systems.
  The CI method is then used to evaluate valence-valence
  correlations.

  First, we  express the all-order equations
   Eqs.~(\ref{v1}-\ref{env}) in terms of matrix elements of $\Sigma_1$ and  $\Sigma_2$ and explicitly
   include the energy dependence. We also need to add an all-order equation for the excitation coefficients
   $\rho_{mnvw}$ to obtain $\Sigma_2$. This equation is equivalent to Eq.~(\ref{v2}) with core index $b$ replaced
   by valence index $w$.

 $\Sigma_1$ and  $\Sigma_2$ are essentially the all-order excitation coefficients $\rho_{mv}$ and $\rho_{mnvw}$:
  \begin{eqnarray*}
  \Sigma_{ma} &=& \rho_{ma} \left( \epsilon_a-\epsilon_m \right) \\
  \Sigma_{mnab} &=& \rho_{mnab} \left( \epsilon_{ab}-\epsilon_{mn} \right) \\
  \Sigma_{mnva} &=& \rho_{mnva} \left(\tilde\epsilon_{v} + \epsilon_{a}-\epsilon_{mn} \right) \\
  \Sigma_{mv} &\equiv&\left(\Sigma_1\right)_{mv}= \rho_{mv} \left( \tilde\epsilon_v-\epsilon_m \right) \\
    \Sigma_{mnvw} &\equiv&\left(\Sigma_2\right)_{mnvw}= \rho_{mnvw} \left(\tilde\epsilon_{v} + \tilde\epsilon_{w}-\epsilon_{mn} \right)
  \end{eqnarray*}

  The quantities  $\Sigma_{ma}$, $\Sigma_{mnab}$, $\Sigma_{mnva}$ are used in the all-order iteration procedure
  but do not explicitly appear in the effective Hamiltonian.
      The core equations for $\rho_{ma}$ and $\rho_{mnab}$ are not modified from the original all-order monovalent code. The excitation coefficients
  $\rho_{ma}$ and $\rho_{mnab}$  are simply multiplied by the appropriate energy differences
  to obtain the terms $\Sigma_{ma}$ and $\Sigma_{mnab}$ needed by other programs.
  Re-writing the other all-order equations in terms of $\Sigma$ and removing terms that will
   be otherwise double-counted by the
 CI part of the calculations, we obtain the following set of equations:
  \begin{eqnarray}
  \label{sigma1}
\Sigma_{mv}&\equiv&\left(\Sigma_1\right)_{mv}= \sum_{nb}\frac{\tilde{g}_{mbvn}\,\Sigma_{nb}}{\epsilon_{bn}+\tilde{\epsilon}_v-\epsilon_v}\hspace{0.3in}\\
&-&\sum_{bcn}\frac{\tilde{g}_{bcvn}\,\Sigma_{mnbc}}{\epsilon_{bc}-\epsilon_{mn}+\tilde{\epsilon}_v-\epsilon_v}
+\sum_{bnr}\frac{\tilde{g}_{mbnr}\,\Sigma_{nrvb}}{\tilde\epsilon_{v}+\epsilon_{b}-\epsilon_{nr}}, \nonumber
\end{eqnarray}
   \begin{eqnarray}
        \label{sigma2-va}
&&\Sigma_{mnvb}= g_{mnvb}+\sum_{cd}\frac{g_{cdvb}\,\Sigma_{mncd}}{\epsilon_{cd}-\epsilon_{mn}+ \tilde{\epsilon}_v-\epsilon_v}\\
&&+\sum_{rs}\frac{g_{mnrs}\,\Sigma_{rsvb}}{\tilde{\epsilon}_v+\epsilon_{b}-\epsilon_{rs}}
-\sum_{c}\frac{g_{cnvb}\,\Sigma_{mc}}{\epsilon_{c}-\epsilon_{m}+\tilde{\epsilon}_v-\epsilon_{v}}\nonumber\\
&&+\sum_{r}\frac{g_{mnvr}\,\Sigma_{rb}}{\epsilon_{b}-\epsilon_{r}+\tilde{\epsilon}_v-\epsilon_{v}}
-\sum_{c}\frac{g_{mcvb}\,\Sigma_{nc}}{\epsilon_{c}-\epsilon_{n}+\tilde{\epsilon}_v-\epsilon_{v}}\nonumber\\
&&+\sum_{cr}\frac{\tilde{g}_{cnrb}\,\Sigma_{mrvc}}{\tilde{\epsilon}_v+\epsilon_{c}-\epsilon_{mr}}
-\sum_{cr}\frac{{g}_{cnrb}\,\Sigma_{rmvc}}{\tilde{\epsilon}_v+\epsilon_{c}-\epsilon_{mr}}\nonumber\\
&&-\sum_{cr}\frac{g_{mcrb}\,\Sigma_{rnvc}}{\tilde{\epsilon}_v+\epsilon_{c}-\epsilon_{nr}}
+\sum_{cr}\frac{\tilde{g}_{mcvr}\,\Sigma_{rncb}}{\epsilon_{cb}-\epsilon_{nr}+\tilde{\epsilon}_v-\epsilon_{v}}\nonumber\\
&&-\sum_{cr}\frac{g_{mcvr}\,\Sigma_{nrcb}}{\epsilon_{cb}-\epsilon_{nr}+\tilde{\epsilon}_v-\epsilon_{v}}
-\sum_{cr}\frac{g_{cnvr}\,\Sigma_{mrcb}}{\epsilon_{cb}-\epsilon_{mr}+\tilde{\epsilon}_v-\epsilon_{v}},\nonumber
\end{eqnarray}

   \begin{eqnarray}
     \label{sigma2-vw}
\Sigma_{mnvw}& \equiv&\left(\Sigma_2\right)_{mnvw}=\\
&&\sum_{cd}\frac{g_{cdvw}\,\Sigma_{mncd}}{\epsilon_{cd}-\epsilon_{mn}+ \tilde{\epsilon}_v-\epsilon_v+ \tilde{\epsilon}_w-\epsilon_w}\nonumber \\
&-&\sum_{c}\frac{g_{cnvw}\,\Sigma_{mc}}{\epsilon_{c}-\epsilon_{m}+\tilde{\epsilon}_v-\epsilon_{v}+\tilde{\epsilon}_w-\epsilon_{w}}\nonumber\\
&-&\sum_{c}\frac{g_{mcvw}\,\Sigma_{nc}}{\epsilon_{c}-\epsilon_{n}+\tilde{\epsilon}_v-\epsilon_{v}+\tilde{\epsilon}_w-\epsilon_{w}}\nonumber\\
&+&\sum_{cr}\frac{\tilde{g}_{cnrw}\,\Sigma_{mrvc}}{\tilde{\epsilon}_v+\epsilon_{c}-\epsilon_{mr}+\tilde{\epsilon}_w-\epsilon_{w}}\nonumber\\
&-&\sum_{cr}\frac{{g}_{cnrw}\,\Sigma_{rmvc}}{\tilde{\epsilon}_v+\epsilon_{c}-\epsilon_{mr}+\tilde{\epsilon}_w-\epsilon_{w}}\nonumber\\
&-&\sum_{cr}\frac{g_{mcrw}\,\Sigma_{rnvc}}{\tilde{\epsilon}_v+\epsilon_{c}-\epsilon_{nr}+\tilde{\epsilon}_w-\epsilon_{w}}\nonumber\\
&+&\sum_{cr}\frac{\tilde{g}_{mcvr}\,\Sigma_{rncw}}{\epsilon_{c}+\tilde{\epsilon}_w-\epsilon_{nr}+\tilde{\epsilon}_v-\epsilon_{v}}\nonumber\\
&-&\sum_{cr}\frac{g_{mcvr}\,\Sigma_{rnwc}}{\epsilon_{c}+\tilde{\epsilon}_w-\epsilon_{nr}+\tilde{\epsilon}_v-\epsilon_{v}}\nonumber\\
&-&\sum_{cr}\frac{g_{cnvr}\,\Sigma_{rmwc}}{\epsilon_{c}+\tilde{\epsilon}_w-\epsilon_{mr}+\tilde{\epsilon}_v-\epsilon_{v}}.\nonumber
\end{eqnarray}
  The energy denominators are now explicitly written out and the energy dependence is introduced
  following the prescription of the
  CI + second-order MBPT approach. Putting  $\tilde\epsilon_{v}=\epsilon_v$  yields the
  original all-order  equations (\ref{v1}) and (\ref{v2}) for monovalent systems up to $\delta E_v$ terms
  on the left-hand side.
  The term containing $\rho_{rv}$ is removed from Eq.~(\ref{sigma2-va}) since it is included in the CI calculation.
  Eq.~(\ref{sigma2-vw}) for $\Sigma_{mnvw}$
  does not have terms that would corresponds to the first, third, and fifth terms
  of Eq.~(\ref{sigma2-va}) for $\Sigma_{mnva}$,
  since these terms are accounted for by the CI as well. Therefore, no iteration is required in the last equation
 since $\Sigma_{mnvw}$ does not appear on its right-hand side; the last equation needs to be evaluated
    only once after all other equations have converged.
    The last equation is also significantly faster to evaluate than similar equations for
     $\Sigma_{mnab}$ and $\Sigma_{mnva}$ since it does not
    contain the term with four indexes over the virtual orbitals (term 3 in Eq.~\ref{sigma2-va}).
    We note that this is the only equation that is not present in any form in the
    all-order code for monovalent systems.

      Below, we outline a step-by-step application of the CI + all-order  method.

  \textbf{Step 1}: A finite basis set is generated in a spherical cavity of radius R. All calculations are
  carried out with same basis set. The second-order MBPT is used to generate $(\Sigma^{(2)}_1)_{vw}$ and
   $(\Sigma^{(2)}_2)_{mnvw}$, where $m,n,v,w$ are any basis set functions. Highly-excited
    orbitals can be omitted without loss of accuracy.  A subset of the basis set
     orbitals is selected for which all-order values of $(\Sigma_1)_{vw}$ and
   $(\Sigma_2)_{mnvw}$ are to be obtained.
 In our calculation this set includes the  first three $ns$, $np_{1/2}$, $np_{3/2}$, $nd_{3/2}$, $nd_{5/2}$ states for each system.
 We found that inclusion of the additional orbitals did not significantly change the results and treating the remaining corrections
 to the effective Hamiltonian in second order should be sufficiently accurate. This issue is addressed in more detail in the next section.

  \textbf{Step 2}: The all-order core $\rho_{ma}$ and $\rho_{mnab}$ excitation
  coefficients are obtained by the iterative solution of the corresponding equations in the
  appropriate potential (for example, $V^{N-2}$ for divalent systems) in the same way as for monovalent systems.
   The core correlation energy is used
  as a convergence parameter and is generally required to converge to 10$^{-5}$ relative accuracy.
  The core excitation coefficients are multiplied by the appropriate denominators as described above to obtain
  $\Sigma_{ma}$ and $\Sigma_{mnab}$ after the iterations are complete.

  \textbf{Step 3}: The core quantities $\Sigma_{ma}$ and $\Sigma_{mnab}$ are used to obtain
   $\Sigma_{mv}$ and $\Sigma_{mnva}$,
   again by an iteration procedure, for a large number of excited $m,n$, and $v$
orbitals. The valence correlation energy for the state $v$ is used
as a convergence parameter.  These steps are carried out in exactly
the same way as our present monovalent all-order calculations with
the omission of the valence-valence diagrams as described above. The
iterations of excitation coefficients result in the summation
of the relevant classes of MBPT terms to all orders. We note that
the term $\Sigma_{mv}$ gives the all-order correction to the
one-body part of the effective Hamiltonian.

     \textbf{Step 4}: The all-order expression for $(\Sigma_2)_{mnvw}$ corrections to the
     effective Hamiltonian are calculated using Eq.~(\ref{sigma2-vw}) with previously stored, fully converged,
     values of
     $\Sigma_{ma}$, $\Sigma_{mnab}$, and $\Sigma_{mnvb}$.

      \textbf{Step 5}:  CI calculations are carried out to generate accurate wave functions
      with the effective Hamiltonian constructed using $\Sigma_1$ and $\Sigma_2$
      obtained in the previous steps.

       \textbf{Step 6}:  The resulting wave functions are used to obtain various matrix elements and
       derived quantities such as PNC amplitudes. In the current CI + MBPT approach \cite{Savukov:2002},
       matrix elements are calculated by replacing ``bare'' matrix elements by the
       ``dressed'' matrix elements using the  random-phase approximations
       (RPA). In this work, we use the same approach. This
       issue will be further discussed in the later section.

       The method described above treats electronic correlation in systems with several valence electrons in a
       significantly more complete way than the CI + MBPT approach owing to the inclusion of the additional classes of
        MBPT terms in
        $\Sigma_1$ and addition of all-order (rather than second-order) correction in $\Sigma_2$.
        We note that our present all-order code is capable of efficiently evaluating the
        large number of the core-valence all-order excitation coefficients needed
        for the implementation of the CI + all-order approach.

 \begin{table*}
\caption{Comparison of the CI, CI+MBPT  and CI+all-order \textit{ab initio} results for the two-electron binding energies of
Mg, Ca, Zn, Sr, Cd, Ba, and Hg with experiment. The energies are given in cm$^{-1}$.
 The relative difference with experimental values is given in the last three columns in \%. %%@
\label{tab1} }
\begin{ruledtabular}
\begin{tabular}{llccccccc}
 \multicolumn{1}{c}{Element}&
\multicolumn{1}{c}{State}&
 \multicolumn{4}{c}{Energies}&
\multicolumn{3}{c}{Differences with experiment (\%)}\\
 \multicolumn{1}{c}{}&
\multicolumn{1}{c}{}&
 \multicolumn{1}{c}{Expt.}&
 \multicolumn{1}{c}{CI}&
\multicolumn{1}{c}{CI+MBPT}&
\multicolumn{1}{c}{CI+all-order}&
\multicolumn{1}{c}{CI}&
\multicolumn{1}{c}{CI+MBPT}&
\multicolumn{1}{c}{CI+all-order}\\
\hline
Mg&  $3s^2 \, ^1S_0$&  182939&  179537&  182717&  182877&  1.86&   0.12&   0.03\\
Ca&  $4s^2 \, ^1S_0$&  145058&  139068&  145985&  145517&  4.13&  -0.64&  -0.32\\
Zn&  $4s^2 \, ^1S_0$&  220662&  204083&  218521&  219442&  7.51&   0.97&   0.55\\
Sr&  $5s^2 \, ^1S_0$&  134896&  127858&  136082&  135322&  5.22&  -0.88&  -0.32\\
Cd&  $5s^2 \, ^1S_0$&  208915&  188884&  210716&  208620&  9.59&  -0.86&   0.14\\
Ba&  $6s^2 \, ^1S_0$&  122721&  114898&  124956&  123363&  6.37&  -1.82&  -0.52\\
Hg&  $6s^2\, ^1S_0$&    235469& 207652& 241152& 236626& 11.81&  -2.41&  -0.49\\
\end{tabular}
\end{ruledtabular}
\end{table*}

 \begin{table*}
\caption{Comparison of the CI, CI+MBPT  and CI+all-order \textit{ab initio} results for the energy levels of
Mg, Ca,  Cd, and Ba with experiment. Two-electron binding energies are given
in the first row for each element, the other values are counted from the ground state
energy. The energies are given in cm$^{-1}$.
 The relative difference with experimental values is given in the last three columns in \%. %%@
\label{tab2} }
\begin{ruledtabular}
\begin{tabular}{llccccccc}
 \multicolumn{1}{c}{Element}&
\multicolumn{1}{c}{State}&
 \multicolumn{4}{c}{Energies}&
\multicolumn{3}{c}{Differences with experiment (\%)}\\
 \multicolumn{2}{c}{}&
 \multicolumn{1}{c}{Expt.}&
 \multicolumn{1}{c}{CI} &
\multicolumn{1}{c}{CI+MBPT}&
\multicolumn{1}{c}{CI+all-order}&
\multicolumn{1}{c}{CI}&
\multicolumn{1}{c}{CI+MBPT}&
\multicolumn{1}{c}{CI+all-order}\\
\hline
Mg &$3s^2 \, ^1S_0$ &  182939 & 179537 & 182717 &  182877 & 1.86 &  0.12 &  0.03\\
&$3s4s \, ^3S_1$ & 41197   & 40409   &41132  &  41175  & 1.91 &  0.16 &  0.05\\
&$3s4s \, ^1S_0$ & 43503   & 42689   &43459  &  43502  & 1.87 &  0.10 &  0.00\\
&$3s3d \, ^1D_2$ & 46403   & 45119   &46318  &  46384  & 2.77 &  0.18 &  0.04\\
&$3s3d \, ^3D_1$ & 47957   & 46972   &47892  &  47936  & 2.05 &  0.14 &  0.04\\
&$3s3d \, ^3D_2$ & 47957   & 46972   &47892  &  47936  & 2.05 &  0.14 &  0.04\\
&$3s3d \, ^3D_3$ & 47957   & 46972   &47892  &  47936  & 2.05 &  0.14 &  0.04\\
&$3s3p \, ^3P_0$ & 21850   & 20906   &21780  &  21833  & 4.32 &  0.32 &  0.08\\
&$3s3p \, ^3P_1$ & 21870   & 20926   &21801  &  21852  & 4.32 &  0.32 &  0.08\\
&$3s3p \, ^3P_2$ & 21911   & 20967   &21844  &  21897  & 4.31 &  0.30 &  0.07\\
&$3s3p \, ^1P_1$ & 35051   & 34488   &35053  &  35065  & 1.61 &  0.00 & -0.04\\[0.5pc]
Ca &$4s^2  \,  ^1S_0$ &145058 & 139068 & 145985 &145517 &  4.13  & -0.64 &  -0.32\\
&$3d4s \,  ^3D_1$ &20335  & 24200  & 19927  & 20335 & -19.00 &  2.01 &   0.00\\
&$3d4s \,  ^3D_2$ &20349  & 24201  & 19949  & 20355 & -18.93 &  1.97 &  -0.03\\
&$3d4s \,  ^3D_3$ &20371  & 24203  & 19982  & 20386 & -18.81 &  1.91 &  -0.07\\
&$3d4s \,  ^1D_2$ &21850  & 23853  & 21620  & 21965 & -9.17  &  1.05 &  -0.53\\
&$4s5s \,  ^3S_1$ &31539  & 30147  & 31765  & 31694 &  4.42  & -0.72 &  -0.49\\
&$4s5s \,  ^1S_0$ &33317  & 31893  & 33552  & 33466 &  4.27  & -0.70 &  -0.45\\
&$4s4p \,  ^3P_0$ &15158  & 13509  & 15474  & 15338 & 10.88  & -2.08 &  -1.19\\
&$4s4p \,  ^3P_1$ &15210  & 13557  & 15528  & 15385 & 10.87  & -2.09 &  -1.15\\
&$4s4p \,  ^3P_2$ &15316  & 13655  & 15638  & 15498 & 10.85  & -2.10 &  -1.19\\
&$4s4p \,  ^1P_1$ &23652  & 23052  & 23771  & 23729 &  2.54  & -0.50 &  -0.32\\[0.5pc]
Cd&$5s^2\,^1S_0$ & 208915 &  188884 & 210716& 208620 &  9.59 &  -0.86 &  0.14\\
&$5s6s\, ^3S_1$ &  51484 &  44027  & 51916  & 51395  & 14.48 &  -0.84 &  0.17\\
&$5s6s\, ^1S_0$ &  53310 &  46153  & 53788  & 53272  & 13.43 &  -0.90 &  0.07\\
&$5s5d\,^1D_2$ &  59220 &  50634  & 59697  & 59015  & 14.50 &  -0.81 &  0.35\\
&$5s5d\, ^3D_1$ &  59486 &  51292  & 59881  & 59259  & 13.77 &  -0.66 &  0.38\\
&$5s5d\, ^3D_2$ &  59498 &  51303  & 59893  & 59271  & 13.77 &  -0.66 &  0.38\\
&$5s5d\, ^3D_3$ &  59516 &  51320  & 59911  & 59291  & 13.77 &  -0.66 &  0.38\\
&$5s5p\, ^3P_0$ &  30114 &  24417  & 30903  & 30141  & 18.92 &  -2.62 & -0.09\\
&$5s5p\, ^3P_1$ &  30656 &  24875  & 31451  & 30646  & 18.86 &  -2.59 &  0.03\\
&$5s5p\, ^3P_2$ &  31827 &  25833  & 32656  & 31838  & 18.83 &  -2.60 & -0.03\\
&$5s5p\, ^1P_1$ &  43692 &  38902  & 43970  & 43607  & 10.96 &  -0.64 &  0.20\\[0.5pc]
Ba&$6s^2 \, ^1S_0 $&  122721 &  114898&   124956 &  123363&   6.37 &  -1.82  & -0.52\\
&$6s5d \, ^3D_1 $&  9034   &  11524 &   9276   &  9249  & -27.57 &  -2.67  & -2.38\\
&$6s5d \, ^3D_2 $&  9216   &  11603 &   9489   &  9441  & -25.91 &  -2.97  & -2.45\\
&$6s5d \, ^3D_3 $&  9597   &  11780 &   9941   &  9840  & -22.75 &  -3.59  & -2.54\\
&$6s5d \, ^1D_2 $&  11395  &  12753 &   11878  &  11727 &  -11.92&   -4.24 & -2.91\\
&$6s6p \, ^3P_0 $&  12266  &   9938 &  13112   & 12556  &  18.98 &  -6.90  & -2.36\\
&$6s6p \, ^3P_1 $&  12637  &   10269&   13484  & 12919  &  18.73 &  -6.70  & -2.23\\
&$6s6p \, ^3P_2 $&  13515  &   11010&   14391  & 13819  &  18.53 &  -6.48  & -2.25\\
&$6s6p \, ^1P_1 $&  18060  &   16908&   18621  & 18292  &   6.38 &  -3.11  & -1.28\\
\end{tabular}
\end{ruledtabular}
\end{table*}

\begin{table}
\caption{Comparison of the CI + all-order results for the energy levels of
Ba with experiment. Two-electron binding energies are given
in the first row, the other values are counted from the ground state.
The one-particle $\tilde\epsilon$ energies for $ns$ and $np$ orbitals
 are set to $\tilde\epsilon_{ns} = -0.48$~a.u. and
$\tilde\epsilon_{np_{1/2,3/2}} = -0.40$~a.u., the energies for the other partial waves are set to
the Dirac-Fock values for the lowest orbital.
 The energies are given in cm$^{-1}$.
 The relative difference with experimental values is given in the last column in \%. %%@
\label{tab3} }
\begin{ruledtabular}
\begin{tabular}{lrrrr}
\multicolumn{1}{l}{State}&
 \multicolumn{1}{c}{Expt.}&
\multicolumn{1}{c}{CI+all-order}&
\multicolumn{1}{c}{$\Delta(\%)$}\\
\hline
$6s^2 \,  ^1S_0$    &122721 &122757  & -0.03\\
$6s5d \,  ^3D_1$    &9034   &  9012  & 0.25\\
$6s5d \,  ^3D_2$    &9216   &  9202  & 0.14\\
$6s5d \,  ^3D_3$    &9597   &  9603  & -0.07\\
$6s5d \,  ^1D_2$    &11395  &  11407 &   -0.11\\
$6s6p \, ^3P_0 $    &12266  &  12235 &   0.26\\
$6s6p \, ^3P_1 $    &12637  &  12602 &   0.27\\
$6s6p \, ^3P_2 $    &13515  &  13491 &   0.18\\
$6s6p \, ^1P_1 $    &18060  &  18120 &   -0.33\\
$5d6p \, ^3F_2 $    &22065  &  22201 &   -0.62\\
$5d6p \,  ^3F_3$    &22947  &  23174 &   -0.99\\
$5d6p \,  ^3F_4$    &23757  &  24005 &   -1.04\\
$\Delta(^3D_1 -~^3P_0)$&3232 &  3223    &0.28\\
\end{tabular}
\end{ruledtabular}
\end{table}

\section{Results and discussions}

        We compare the results of our CI, CI + MBPT, and
CI + all-order \textit{ab initio} calculations for the two-electron binding energies of
Mg, Ca, Zn, Sr, Cd, Ba, and Hg with experiment in Table~\ref{tab1}.
Results for the  energies of Mg, Ca, Cd, and Ba, counted from the ground state, are compared with experiment
in Table~\ref{tab2}. The same designations are used in both tables.
The energy values are given in cm$^{-1}$. Relative differences of our results
with experiment are given in the last
three columns of Tables \ref{tab1} and \ref{tab2} to illustrate the accuracy of each approach.

The same parameters and basis set are used in all three calculations for each system.
The finite basis set of 245 orbitals that include
$l=0\dots5$ partial waves is formed in the spherical cavity with a 50~a.u. radius.
 The CI calculation
includes only valence shell excitations as described above, i.e. CI
calculation is carried out in the same way for all three cases.
All  summations over the excited states in the second-order and the all-order
calculations are always carried out over the entire basis set.

A sufficiently large number of the effective Hamiltonian matrix elements
are modified in the CI + second-order MBPT calculation.
 There is no need to include corrections to the entire Hamiltonian
as the corrections from the remaining terms are negligible as described below.
The CI+all-order calculations include replacement of the most important $\Sigma_1$ and $\Sigma_2$  terms
by their all-order values. The remaining corrections from the effective Hamiltonian retain their second-order values as described above.
We find that it is sufficient to carry out all-order calculations
for the first three $ns$, $np_{1/2}$, $np_{3/2}$, $nd_{3/2}$, $nd_{5/2}$ states and modify the
corresponding $\Sigma_1$ and $\Sigma_2$. For example, indexes $m,n,v,w$ in $(\Sigma_1)_{mv}$ and $(\Sigma_2)_{mnvw}$ in Ca calculation include $4s$, $5s$, $6s$, $4p_{1/2}$, $5p_{1/2}$, $6p_{1/2}$, $4p_{3/2}$, $5p_{3/2}$, $6p_{3/2}$,  $3d_{3/2}$, $4d_{3/2}$, $5d_{3/2}$,
$3d_{5/2}$, $4d_{5/2}$, and $5d_{5/2}$ states.
To test that the above number of the corrected Hamiltonian matrix elements is sufficient, we have carried out the following test in Ca:

 (1) the number of second-order
$\Sigma_1$ and $\Sigma_2$ matrix elements included
 was increased from 535 to 878 and from  4~879~832 to 19~236~743, respectively;

(2) the number of second-order  $\Sigma_1$ and $\Sigma_2$ matrix elements replaced by
all-order values was increased from 168 to 305 and from  592~634 to 2~898~122, respectively,
by including the $7s$, $7p_{1/2}$, $7p_{3/2}$, $6d_{3/2}$,  $6d_{5/2}$, $4f_{5/2}$, $4f_{7/2}$, $5f_{5/2}$, $5f_{7/2}$ states into the $m,n,v,w$ index set.

The ionization potential and most of the Ca levels that we
considered shifted by less than 1~cm$^{-1}$. The largest changes,
observed for the $4s3d$ levels, were still very small,  0.1~\%. This
was expected since the $nd$ levels are known to be the most affected
by the partial wave restrictions. We note that the energies
$\tilde\epsilon_v$ and $\tilde\epsilon_w$ in
Eqs.~(\ref{sigma1}-\ref{sigma2-vw}) were set in the present
calculation to the corresponding Dirac-Fock values for the lowest
state for each partial wave. The second-order calculations were
carried out in the same way. Therefore, the results listed in
Tables~\ref{tab1} and \ref{tab2} are completely \textit{ab initio}.

We find that the all-order ionization potential results are in significantly better agreement with experiment
in comparison with the CI+MBPT values even in the case of Mg where the agreement with experiment is already excellent in the CI+MBPT approach.
We also find almost no deterioration
in the accuracy of the two-electron binding energies from Ca to Hg; the all-order
 method reduces the differences with experiment by about factor of three in comparison with the second-order data.
 Similar improvements are observed for most of the excited states listed in Table~\ref{tab2}
  with the exception of $5d6s$ states of Ba. The accuracy of the SD all-order approach is expected to
  be lower for Ba since the SD method omits certain parts of the third-order energy correction
  associated with valence triple excitations (last term in Eq.~(\ref{eq1})).
  This contribution was found to increase significantly  for heavier alkalis \cite{SDpT}.
  The problem is corrected in the  SD all-order method by explicitly adding the missing part of the third-order correction.
  Within our approach, this issue may be treated in an \textit{ab initio} way by adding the valence triple excitation
  terms perturbatively to the all-order as was done for the monovalent systems in Ref.~\cite{SDpT}
  and removing terms that are accounted for by the CI.
  We note that factor of three improvement in the relative differences with experiment  is still observed
  for the $6s6p$ energies calculated with the all-order method.

We observe that essentially all of the states in Ba listed in Table~\ref{tab2} are shifted by the same relative amount in the
all-order approach, unlike the case of CI + MBPT. In this case, it is possible to carry out another
calculation
with different $\tilde\epsilon_v$ and $\tilde\epsilon_w$ resulting in final energies in very close
agreement with experiment in order to get an improved representation of the wave functions for subsequent evaluation of atomic properties.
The results of such a calculation for Ba
are given in Table~\ref{tab3}. The one-particle $\tilde\epsilon$ energies for $ns$ and $np$ orbitals in this calculation
 are set to $\tilde\epsilon_{ns} = -0.48$~a.u. and
$\tilde\epsilon_{np_{1/2,3/2}} = -0.40$~a.u., the energies for the other partial waves are set to
the Dirac-Fock values for the lowest orbital just as in the previous calculation.
The results of this calculation are in excellent agreement with experiment. Moreover,
the $\Delta(^3D_1 -~^3P_0)$ energy difference which is very difficult to accurately calculate
agrees with experiment  to 11~cm$^{-1}$.

Our calculations have yielded accurate wave functions for subsequent evaluation of the atomic properties.
Matrix elements of one-body operators such as E1, E2, hyperfine, parity-violation, etc.,
 are evaluated in the framework of the CI + MBPT approach in the RPA approximation  \cite{Dzuba:1996} (as described in Step 6 of the previous section),
 sometimes with subsequent addition of the dominant normalization and structure radiation terms.
 In the CI + all-order method, we can use exactly the same method to evaluate matrix elements.
 Our preliminary calculations of the $^3P_0$ polarizability values for Ca and Sr  indicate better agreement of the
 CI+all-order \textit{ab initio} results with recommended values from Ref.~\cite{POL-Andrei} in comparison with the CI+MBPT approach.
  The complete implementation of the all-order approach would require addition of the
all-order corrections to the matrix elements beyond the modification of the wave function that is the subject of the present paper.
This approach will implicitly include dominant normalization, structure radiation, and other corrections
to all orders.
 In further work, we plan to replace the RPA matrix elements  by all-order
 counterparts that are linear or quadratic  functions of the excitation coefficients \cite{SDpT}. The
 terms that are accounted by the CI will have to be removed to avoid double counting.
 The ability to conduct calculations in various approximations will also allow one to carry out the evaluation of
 uncertainties of atomic properties needed for many application, such as calculation of BBR shifts and the study of fundamental symmetries.

\section{Conclusion}
We have developed a theoretical method combining the all-order approach currently used in precision calculations
of properties of monovalent atoms with the configuration-interaction approach that is
applicable for many-electron systems. This approach has been tested on the calculation of energy levels of divalent systems from Mg to Hg.
We have demonstrated an improvement of at least a factor of three in agreement with experimental values for the two-electron binding energies
and most excited state energies in comparison with the CI + MBPT method. Further work on this method will include addition of the all-order
terms beyond the RPA in the treatment of the transition amplitudes
and other matrix elements for precision calculation of atomic properties of systems with few valence electrons.

\acknowledgments

This work
was supported in part by US National Science Foundation Grant  No.\
PHY-07-58088 and by the RFBR grant No. 08-02-00460. MGK thanks University of Delaware
for hospitality.

% \bibliography{paperCI}

\begin{thebibliography}{39}
\expandafter\ifx\csname natexlab\endcsname\relax\def\natexlab#1{#1}\fi
\expandafter\ifx\csname bibnamefont\endcsname\relax
  \def\bibnamefont#1{#1}\fi
\expandafter\ifx\csname bibfnamefont\endcsname\relax
  \def\bibfnamefont#1{#1}\fi
\expandafter\ifx\csname citenamefont\endcsname\relax
  \def\citenamefont#1{#1}\fi
\expandafter\ifx\csname url\endcsname\relax
  \def\url#1{\texttt{#1}}\fi
\expandafter\ifx\csname urlprefix\endcsname\relax\def\urlprefix{URL }\fi
\providecommand{\bibinfo}[2]{#2}
\providecommand{\eprint}[2][]{\url{#2}}

\bibitem[{\citenamefont{Blundell et~al.}(1991)\citenamefont{Blundell, Johnson,
  and Sapirstein}}]{SD}
\bibinfo{author}{\bibfnamefont{S.~A.} \bibnamefont{Blundell}},
  \bibinfo{author}{\bibfnamefont{W.~R.} \bibnamefont{Johnson}},
  \bibnamefont{and}
  \bibinfo{author}{\bibfnamefont{J.}~\bibnamefont{Sapirstein}},
  \bibinfo{journal}{Phys. Rev. A} \textbf{\bibinfo{volume}{43}},
  \bibinfo{pages}{3407} (\bibinfo{year}{1991}).

\bibitem[{\citenamefont{Safronova et~al.}(1999)\citenamefont{Safronova,
  Johnson, and Derevianko}}]{SDpT}
\bibinfo{author}{\bibfnamefont{M.~S.} \bibnamefont{Safronova}},
  \bibinfo{author}{\bibfnamefont{W.~R.} \bibnamefont{Johnson}},
  \bibnamefont{and}
  \bibinfo{author}{\bibfnamefont{A.}~\bibnamefont{Derevianko}},
  \bibinfo{journal}{Phys. Rev. A} \textbf{\bibinfo{volume}{60}},
  \bibinfo{pages}{4476} (\bibinfo{year}{1999}).

\bibitem[{\citenamefont{Safronova and Johnson}(2007)}]{review07}
\bibinfo{author}{\bibfnamefont{M.~S.} \bibnamefont{Safronova}}
  \bibnamefont{and} \bibinfo{author}{\bibfnamefont{W.~R.}
  \bibnamefont{Johnson}}, \bibinfo{journal}{Adv. At. Mol., Opt. Phys.}
  \textbf{\bibinfo{volume}{55}}, \bibinfo{pages}{191} (\bibinfo{year}{2007}).

\bibitem[{\citenamefont{Vasilyev et~al.}(2002)\citenamefont{Vasilyev, Savukov,
  Safronova, and Berry}}]{pnc-cs}
\bibinfo{author}{\bibfnamefont{A.~A.} \bibnamefont{Vasilyev}},
  \bibinfo{author}{\bibfnamefont{I.~M.} \bibnamefont{Savukov}},
  \bibinfo{author}{\bibfnamefont{M.~S.} \bibnamefont{Safronova}},
  \bibnamefont{and} \bibinfo{author}{\bibfnamefont{H.~G.} \bibnamefont{Berry}},
  \bibinfo{journal}{Phys. Rev. A} \textbf{\bibinfo{volume}{66}},
  \bibinfo{pages}{020101} (\bibinfo{year}{2002}).

\bibitem[{\citenamefont{Safronova and Johnson}(2000)}]{pnc-fr}
\bibinfo{author}{\bibfnamefont{M.~S.} \bibnamefont{Safronova}}
  \bibnamefont{and} \bibinfo{author}{\bibfnamefont{W.~R.}
  \bibnamefont{Johnson}}, \bibinfo{journal}{Phys. Rev. A}
  \textbf{\bibinfo{volume}{62}}, \bibinfo{pages}{022112}
  (\bibinfo{year}{2000}).

\bibitem[{\citenamefont{Pal et~al.}()\citenamefont{Pal, Jiang, Safronova,
  Johnson, and Safronova}}]{pnc-ra}
\bibinfo{author}{\bibfnamefont{R.}~\bibnamefont{Pal}},
  \bibinfo{author}{\bibfnamefont{D.}~\bibnamefont{Jiang}},
  \bibinfo{author}{\bibfnamefont{M.~S.} \bibnamefont{Safronova}},
  \bibinfo{author}{\bibfnamefont{W.~R.} \bibnamefont{Johnson}},
  \bibnamefont{and} \bibinfo{author}{\bibfnamefont{U.~I.}
  \bibnamefont{Safronova}}, \bibinfo{note}{arXiv:0901.4195 (2009)}.

\bibitem[{\citenamefont{Arora et~al.}(2007{\natexlab{a}})\citenamefont{Arora,
  Safronova, and Clark}}]{magic}
\bibinfo{author}{\bibfnamefont{B.}~\bibnamefont{Arora}},
  \bibinfo{author}{\bibfnamefont{M.~S.} \bibnamefont{Safronova}},
  \bibnamefont{and} \bibinfo{author}{\bibfnamefont{C.~W.} \bibnamefont{Clark}},
  \bibinfo{journal}{Phys. Rev. A} \textbf{\bibinfo{volume}{76}},
  \bibinfo{pages}{052509} (\bibinfo{year}{2007}{\natexlab{a}}).

\bibitem[{\citenamefont{Arora et~al.}(2007{\natexlab{b}})\citenamefont{Arora,
  Safronova, , and Clark}}]{BBR}
\bibinfo{author}{\bibfnamefont{B.}~\bibnamefont{Arora}},
  \bibinfo{author}{\bibfnamefont{M.~S.} \bibnamefont{Safronova}}, ,
  \bibnamefont{and} \bibinfo{author}{\bibfnamefont{C.~W.} \bibnamefont{Clark}},
  \bibinfo{journal}{Phys. Rev. A} \textbf{\bibinfo{volume}{76}},
  \bibinfo{pages}{064501} (\bibinfo{year}{2007}{\natexlab{b}}).

\bibitem[{\citenamefont{Jiang et~al.}(2008)\citenamefont{Jiang, Arora, and
  Safronova}}]{quad}
\bibinfo{author}{\bibfnamefont{D.}~\bibnamefont{Jiang}},
  \bibinfo{author}{\bibfnamefont{B.}~\bibnamefont{Arora}}, \bibnamefont{and}
  \bibinfo{author}{\bibfnamefont{M.~S.} \bibnamefont{Safronova}},
  \bibinfo{journal}{Phys. Rev. A} \textbf{\bibinfo{volume}{78}},
  \bibinfo{pages}{022514} (\bibinfo{year}{2008}).

\bibitem[{\citenamefont{Dzuba et~al.}(1996)\citenamefont{Dzuba, Flambaum, and
  Kozlov}}]{Dzuba:1996}
\bibinfo{author}{\bibfnamefont{V.~A.} \bibnamefont{Dzuba}},
  \bibinfo{author}{\bibfnamefont{V.~V.} \bibnamefont{Flambaum}},
  \bibnamefont{and} \bibinfo{author}{\bibfnamefont{M.~G.}
  \bibnamefont{Kozlov}}, \bibinfo{journal}{Phys.\ Rev.\ A}
  \textbf{\bibinfo{volume}{54}}, \bibinfo{pages}{3948} (\bibinfo{year}{1996}).

\bibitem[{\citenamefont{Porsev et~al.}(2001)\citenamefont{Porsev, Kozlov,
  Rakhlina, and Derevianko}}]{Porsev:2001}
\bibinfo{author}{\bibfnamefont{S.~G.} \bibnamefont{Porsev}},
  \bibinfo{author}{\bibfnamefont{M.~G.} \bibnamefont{Kozlov}},
  \bibinfo{author}{\bibfnamefont{Y.~G.} \bibnamefont{Rakhlina}},
  \bibnamefont{and}
  \bibinfo{author}{\bibfnamefont{A.}~\bibnamefont{Derevianko}},
  \bibinfo{journal}{Phys.\ Rev.\ A} \textbf{\bibinfo{volume}{64}},
  \bibinfo{pages}{012508} (\bibinfo{year}{2001}).

\bibitem[{\citenamefont{Kozlov and Porsev}(1999)}]{Kozlov:1999}
\bibinfo{author}{\bibfnamefont{M.~G.} \bibnamefont{Kozlov}} \bibnamefont{and}
  \bibinfo{author}{\bibfnamefont{S.~G.} \bibnamefont{Porsev}},
  \bibinfo{journal}{Eur.\ Phys.\ J. D} \textbf{\bibinfo{volume}{5}},
  \bibinfo{pages}{59} (\bibinfo{year}{1999}).

\bibitem[{\citenamefont{Kozlov and Porsev}(1997)}]{Kozlov:1997}
\bibinfo{author}{\bibfnamefont{M.~G.} \bibnamefont{Kozlov}} \bibnamefont{and}
  \bibinfo{author}{\bibfnamefont{S.~G.} \bibnamefont{Porsev}},
  \bibinfo{journal}{J.\ Expt.\ Theor. Phys.} \textbf{\bibinfo{volume}{84}},
  \bibinfo{pages}{461} (\bibinfo{year}{1997}).

\bibitem[{\citenamefont{Dzuba and Flambaum}(2007{\natexlab{a}})}]{Dzuba:2007}
\bibinfo{author}{\bibfnamefont{V.~A.} \bibnamefont{Dzuba}} \bibnamefont{and}
  \bibinfo{author}{\bibfnamefont{V.~V.} \bibnamefont{Flambaum}},
  \bibinfo{journal}{Phys. Rev. A} \textbf{\bibinfo{volume}{75}},
  \bibinfo{pages}{052504} (\bibinfo{year}{2007}{\natexlab{a}}).

\bibitem[{\citenamefont{Dzuba and Flambaum}(2007{\natexlab{b}})}]{Dzuba:2007a}
\bibinfo{author}{\bibfnamefont{V.~A.} \bibnamefont{Dzuba}} \bibnamefont{and}
  \bibinfo{author}{\bibfnamefont{V.~V.} \bibnamefont{Flambaum}},
  \bibinfo{journal}{J. Phys. B} \textbf{\bibinfo{volume}{40}},
  \bibinfo{pages}{227} (\bibinfo{year}{2007}{\natexlab{b}}).

\bibitem[{\citenamefont{Savukov and Johnson}(2002)}]{Savukov:2002}
\bibinfo{author}{\bibfnamefont{I.~M.} \bibnamefont{Savukov}} \bibnamefont{and}
  \bibinfo{author}{\bibfnamefont{W.~R.} \bibnamefont{Johnson}},
  \bibinfo{journal}{Phys.\ Rev.\ A} \textbf{\bibinfo{volume}{65}},
  \bibinfo{pages}{042503} (\bibinfo{year}{2002}).

\bibitem[{\citenamefont{Savukov et~al.}(2002)\citenamefont{Savukov, Johnson,
  and Berry}}]{Savukov:2002a}
\bibinfo{author}{\bibfnamefont{I.~M.} \bibnamefont{Savukov}},
  \bibinfo{author}{\bibfnamefont{W.~R.} \bibnamefont{Johnson}},
  \bibnamefont{and} \bibinfo{author}{\bibfnamefont{H.~G.} \bibnamefont{Berry}},
  \bibinfo{journal}{Phys.\ Rev.\ A} \textbf{\bibinfo{volume}{66}},
  \bibinfo{pages}{052501} (\bibinfo{year}{2002}).

\bibitem[{\citenamefont{Savukov}(2003)}]{Savukov:2003}
\bibinfo{author}{\bibfnamefont{I.~M.} \bibnamefont{Savukov}},
  \bibinfo{journal}{J.\ Phys.\ B} \textbf{\bibinfo{volume}{36}},
  \bibinfo{pages}{4789} (\bibinfo{year}{2003}).

\bibitem[{\citenamefont{Dinh et~al.}(2008)\citenamefont{Dinh, Dzuba, and
  Flambaum}}]{DF1}
\bibinfo{author}{\bibfnamefont{T.~H.} \bibnamefont{Dinh}},
  \bibinfo{author}{\bibfnamefont{V.~A.} \bibnamefont{Dzuba}}, \bibnamefont{and}
  \bibinfo{author}{\bibfnamefont{V.~V.} \bibnamefont{Flambaum}},
  \bibinfo{journal}{Phys. Rev. A} \textbf{\bibinfo{volume}{78}},
  \bibinfo{pages}{062502} (\bibinfo{year}{2008}).

\bibitem[{\citenamefont{Dzuba and Flambaum}(2008)}]{DF2}
\bibinfo{author}{\bibfnamefont{V.~A.} \bibnamefont{Dzuba}} \bibnamefont{and}
  \bibinfo{author}{\bibfnamefont{V.~V.} \bibnamefont{Flambaum}},
  \bibinfo{journal}{Phys. Rev. A} \textbf{\bibinfo{volume}{77}},
  \bibinfo{pages}{012515} (\bibinfo{year}{2008}).

\bibitem[{\citenamefont{Kozlov et~al.}(2001)\citenamefont{Kozlov, Porsev, and
  Johnson}}]{KPJ:01}
\bibinfo{author}{\bibfnamefont{M.~G.} \bibnamefont{Kozlov}},
  \bibinfo{author}{\bibfnamefont{S.~G.} \bibnamefont{Porsev}},
  \bibnamefont{and} \bibinfo{author}{\bibfnamefont{W.~R.}
  \bibnamefont{Johnson}}, \bibinfo{journal}{Phys.\ Rev.\ A}
  \textbf{\bibinfo{volume}{64}}, \bibinfo{pages}{052107}
  (\bibinfo{year}{2001}).

\bibitem[{\citenamefont{Porsev et~al.}(2000)\citenamefont{Porsev, Kozlov, and
  Rakhlina}}]{PKR00}
\bibinfo{author}{\bibfnamefont{S.~G.} \bibnamefont{Porsev}},
  \bibinfo{author}{\bibfnamefont{M.~G.} \bibnamefont{Kozlov}},
  \bibnamefont{and} \bibinfo{author}{\bibfnamefont{Y.~G.}
  \bibnamefont{Rakhlina}}, \bibinfo{journal}{Hyperfine Interactions}
  \textbf{\bibinfo{volume}{127}}, \bibinfo{pages}{395} (\bibinfo{year}{2000}).

\bibitem[{\citenamefont{Dzuba and Ginges}(2006)}]{Dzuba:2006}
\bibinfo{author}{\bibfnamefont{V.~A.} \bibnamefont{Dzuba}} \bibnamefont{and}
  \bibinfo{author}{\bibfnamefont{J.~S.~M.} \bibnamefont{Ginges}},
  \bibinfo{journal}{Phys. Rev. A} \textbf{\bibinfo{volume}{73}},
  \bibinfo{pages}{032503} (\bibinfo{year}{2006}).

\bibitem[{\citenamefont{Kozlov}(2004)}]{Kozlov:all}
\bibinfo{author}{\bibfnamefont{M.~G.} \bibnamefont{Kozlov}},
  \bibinfo{journal}{Int. J. Q. Chem.} \textbf{\bibinfo{volume}{100}},
  \bibinfo{pages}{336} (\bibinfo{year}{2004}).

\bibitem[{\citenamefont{Dzuba and Flambaum}(2007{\natexlab{c}})}]{DF:2007}
\bibinfo{author}{\bibfnamefont{V.~A.} \bibnamefont{Dzuba}} \bibnamefont{and}
  \bibinfo{author}{\bibfnamefont{V.~V.} \bibnamefont{Flambaum}},
  \bibinfo{journal}{Phys. Rev. A} \textbf{\bibinfo{volume}{75}},
  \bibinfo{pages}{052504} (\bibinfo{year}{2007}{\natexlab{c}}).

\bibitem[{\citenamefont{Brown and Ravenhall}(1951)}]{BR:51}
\bibinfo{author}{\bibfnamefont{G.~E.} \bibnamefont{Brown}} \bibnamefont{and}
  \bibinfo{author}{\bibfnamefont{D.~G.} \bibnamefont{Ravenhall}},
  \bibinfo{journal}{Proc.\ Roy.\ Soc.\ A} \textbf{\bibinfo{volume}{208}},
  \bibinfo{pages}{552} (\bibinfo{year}{1951}).

\bibitem[{\citenamefont{Coester and K{\"{u}}mmel}(1960)}]{CK:60}
\bibinfo{author}{\bibfnamefont{F.}~\bibnamefont{Coester}} \bibnamefont{and}
  \bibinfo{author}{\bibfnamefont{H.}~\bibnamefont{K{\"{u}}mmel}},
  \bibinfo{journal}{Nucl.\ Phys.} \textbf{\bibinfo{volume}{17}},
  \bibinfo{pages}{477} (\bibinfo{year}{1960}).

\bibitem[{\citenamefont{Pal et~al.}(2007)\citenamefont{Pal, Safronova, Johnson,
  Derevianko, and Porsev}}]{NL}
\bibinfo{author}{\bibfnamefont{R.}~\bibnamefont{Pal}},
  \bibinfo{author}{\bibfnamefont{M.~S.} \bibnamefont{Safronova}},
  \bibinfo{author}{\bibfnamefont{W.~R.} \bibnamefont{Johnson}},
  \bibinfo{author}{\bibfnamefont{A.}~\bibnamefont{Derevianko}},
  \bibnamefont{and} \bibinfo{author}{\bibfnamefont{S.~G.}
  \bibnamefont{Porsev}}, \bibinfo{journal}{Phys. Rev. A}
  \textbf{\bibinfo{volume}{75}}, \bibinfo{pages}{042515}
  (\bibinfo{year}{2007}).

\bibitem[{\citenamefont{Porsev and Derevianko}(2006{\natexlab{a}})}]{AS-1}
\bibinfo{author}{\bibfnamefont{S.~G.} \bibnamefont{Porsev}} \bibnamefont{and}
  \bibinfo{author}{\bibfnamefont{A.}~\bibnamefont{Derevianko}},
  \bibinfo{journal}{Phys.\ Rev.\ A} \textbf{\bibinfo{volume}{73}},
  \bibinfo{pages}{012501} (\bibinfo{year}{2006}{\natexlab{a}}).

\bibitem[{\citenamefont{Derevianko and Porsev}(2007)}]{Andrei:Cs}
\bibinfo{author}{\bibfnamefont{A.}~\bibnamefont{Derevianko}} \bibnamefont{and}
  \bibinfo{author}{\bibfnamefont{S.~G.} \bibnamefont{Porsev}},
  \bibinfo{journal}{Eur. Phys. J. A} \textbf{\bibinfo{volume}{32}},
  \bibinfo{pages}{517} (\bibinfo{year}{2007}).

\bibitem[{\citenamefont{Borschevsky et~al.}(2007)\citenamefont{Borschevsky,
  Eliav, Vilkas, Ishikawa, and Kaldor}}]{CCSD-1}
\bibinfo{author}{\bibfnamefont{A.}~\bibnamefont{Borschevsky}},
  \bibinfo{author}{\bibfnamefont{E.}~\bibnamefont{Eliav}},
  \bibinfo{author}{\bibfnamefont{M.~J.} \bibnamefont{Vilkas}},
  \bibinfo{author}{\bibfnamefont{Y.}~\bibnamefont{Ishikawa}}, \bibnamefont{and}
  \bibinfo{author}{\bibfnamefont{U.}~\bibnamefont{Kaldor}},
  \bibinfo{journal}{Phys. Rev. A} \textbf{\bibinfo{volume}{75}},
  \bibinfo{pages}{042514} (\bibinfo{year}{2007}).

\bibitem[{\citenamefont{Landau et~al.}(2000)\citenamefont{Landau, Eliav,
  Ishikawa, and Kaldor}}]{CCSD-2}
\bibinfo{author}{\bibfnamefont{A.}~\bibnamefont{Landau}},
  \bibinfo{author}{\bibfnamefont{E.}~\bibnamefont{Eliav}},
  \bibinfo{author}{\bibfnamefont{Y.}~\bibnamefont{Ishikawa}}, \bibnamefont{and}
  \bibinfo{author}{\bibfnamefont{U.}~\bibnamefont{Kaldor}},
  \bibinfo{journal}{J. Chem. Phys.} \textbf{\bibinfo{volume}{113}},
  \bibinfo{pages}{9905} (\bibinfo{year}{2000}).

\bibitem[{\citenamefont{Eliav et~al.}(1997)\citenamefont{Eliav, Ishikawa,
  Pyykk\"{o}, and Kaldor}}]{CCSD-3}
\bibinfo{author}{\bibfnamefont{E.}~\bibnamefont{Eliav}},
  \bibinfo{author}{\bibfnamefont{Y.}~\bibnamefont{Ishikawa}},
  \bibinfo{author}{\bibfnamefont{P.}~\bibnamefont{Pyykk\"{o}}},
  \bibnamefont{and} \bibinfo{author}{\bibfnamefont{U.}~\bibnamefont{Kaldor}},
  \bibinfo{journal}{Phys. Rev. A} \textbf{\bibinfo{volume}{56}},
  \bibinfo{pages}{4532} (\bibinfo{year}{1997}).

\bibitem[{\citenamefont{Eliav et~al.}(1996)\citenamefont{Eliav, Kaldor,
  Ishikawa, Seth, and Pyykk\"{o}}}]{CCSD-4}
\bibinfo{author}{\bibfnamefont{E.}~\bibnamefont{Eliav}},
  \bibinfo{author}{\bibfnamefont{U.}~\bibnamefont{Kaldor}},
  \bibinfo{author}{\bibfnamefont{Y.}~\bibnamefont{Ishikawa}},
  \bibinfo{author}{\bibfnamefont{M.}~\bibnamefont{Seth}}, \bibnamefont{and}
  \bibinfo{author}{\bibfnamefont{P.}~\bibnamefont{Pyykk\"{o}}},
  \bibinfo{journal}{Phys. Rev. A} \textbf{\bibinfo{volume}{53}},
  \bibinfo{pages}{3926} (\bibinfo{year}{1996}).

\bibitem[{\citenamefont{Eliav et~al.}(1995)\citenamefont{Eliav, Kaldor, and
  Ishikawa}}]{CCSD-5}
\bibinfo{author}{\bibfnamefont{E.}~\bibnamefont{Eliav}},
  \bibinfo{author}{\bibfnamefont{U.}~\bibnamefont{Kaldor}}, \bibnamefont{and}
  \bibinfo{author}{\bibfnamefont{Y.}~\bibnamefont{Ishikawa}},
  \bibinfo{journal}{Phys. Rev. A} \textbf{\bibinfo{volume}{52}},
  \bibinfo{pages}{291} (\bibinfo{year}{1995}).

\bibitem[{\citenamefont{{Yakobi} et~al.}(2007)\citenamefont{{Yakobi}, {Eliav},
  and {Kaldor}}}]{YEK07}
\bibinfo{author}{\bibfnamefont{H.}~\bibnamefont{{Yakobi}}},
  \bibinfo{author}{\bibfnamefont{E.}~\bibnamefont{{Eliav}}}, \bibnamefont{and}
  \bibinfo{author}{\bibfnamefont{U.}~\bibnamefont{{Kaldor}}},
  \bibinfo{journal}{J. Chem. Phys.} \textbf{\bibinfo{volume}{126}},
  \bibinfo{pages}{184305} (\bibinfo{year}{2007}).

\bibitem[{\citenamefont{Bratsev et~al.}(1977)\citenamefont{Bratsev, Deyneka,
  and Tupitsyn}}]{BDT77}
\bibinfo{author}{\bibfnamefont{V.~F.} \bibnamefont{Bratsev}},
  \bibinfo{author}{\bibfnamefont{G.~B.} \bibnamefont{Deyneka}},
  \bibnamefont{and} \bibinfo{author}{\bibfnamefont{I.~I.}
  \bibnamefont{Tupitsyn}}, \bibinfo{journal}{Bull. Acad. Sci. USSR, Phys. Ser.}
  \textbf{\bibinfo{volume}{41}}, \bibinfo{pages}{173} (\bibinfo{year}{1977}).

\bibitem[{\citenamefont{Kotochigova and Tupitsyn}(1987)}]{KT87}
\bibinfo{author}{\bibfnamefont{S.~A.} \bibnamefont{Kotochigova}}
  \bibnamefont{and} \bibinfo{author}{\bibfnamefont{I.~I.}
  \bibnamefont{Tupitsyn}}, \bibinfo{journal}{J. Phys. B}
  \textbf{\bibinfo{volume}{20}}, \bibinfo{pages}{4759} (\bibinfo{year}{1987}).

\bibitem[{\citenamefont{Porsev and
  Derevianko}(2006{\natexlab{b}})}]{POL-Andrei}
\bibinfo{author}{\bibfnamefont{S.~G.} \bibnamefont{Porsev}} \bibnamefont{and}
  \bibinfo{author}{\bibfnamefont{A.}~\bibnamefont{Derevianko}},
  \bibinfo{journal}{Phys. Rev. A} \textbf{\bibinfo{volume}{74}},
  \bibinfo{pages}{020502(R)} (\bibinfo{year}{2006}{\natexlab{b}}).

\end{thebibliography}

  \end{document}